\documentclass[%
 aip,
 amsmath,amssymb,
 reprint,%
nofootinbib,%
]{revtex4-1}

\usepackage{graphicx}% Include figure files
\usepackage{dcolumn}% Align table columns on decimal point
\usepackage{bm}% bold math
\usepackage{xcolor}
\usepackage{graphicx}
\graphicspath{{Figures/}}
\usepackage[utf8]{inputenc}
\usepackage[T1]{fontenc}
\usepackage{mathptmx}
\usepackage{etoolbox}
\usepackage{epsfig}
\makeatletter
\def\@email#1#2{%
 \endgroup
 \patchcmd{\titleblock@produce}
  {\frontmatter@RRAPformat}
  {\frontmatter@RRAPformat{\produce@RRAP{*#1\href{mailto:#2}{#2}}}\frontmatter@RRAPformat}
  {}{}
}%
\makeatother

\begin{document}

\preprint{AIP/123-QED}

\title[nonlinear acoustic metasurface]{Anomalous wavefront control via nonlinear acoustic metasurface through second-harmonic tailoring and demultiplexing}
% Force line breaks with \\
\author{Zhenkun Lin}

\author{Yuning Zhang}

\author{K. W. Wang}

\author{Serife Tol$^\ast$}
 \email{stol@umich.edu.}
\affiliation{ 
Department of Mechanical Engineering, University of Michigan, Ann Arbor, MI, USA, 48109-2125
}%

\date{\today}% It is always \today, today,

\begin{abstract}
We propose a nonlinear acoustic metasurface concept by exploiting the nonlinearity of the locally resonant unit cells formed by curved beams.~The analytical model is established to explore the nonlinear phenomenon, specifically the second-harmonic generation (SHG) of the acoustic waveguide and validated through numerical and experimental studies. Novel nonlinear acoustic metasurfaces are developed to demultiplex different frequency components and achieve anomalous wavefront control of SHG in the transmitted region. To this end, we demonstrate wave steering, wave focusing, and self-bending propagation. Our results show that the proposed nonlinear metasurface provides an effective and efficient platform to achieve significant SHG, and separate different harmonic components for wavefront control of individual harmonics.~Overall, this study offers new avenues to harness nonlinear effects for acoustic wavefront tailoring and develops new potential toward advanced technologies to manipulate acoustic waves.

\end{abstract}

\maketitle
%\section{Introduction}
Engineered material systems have been widely used in controlling acoustic/elastic waves due to their unique and intriguing properties such as bandgap and ability to slow the wave speed,
%effective negative parameters~\cite{hussein2014dynamics,ma2016acoustic} %These engineered structural and material systems exhibit intriguing system dynamic properties, 
which leads to diverse applications, including vibration suppression systems~\cite{barnhart2019experimental,lin2021piezoelectric}, energy harvesting devices~\cite{tol2017phononic,hu2021acoustic}, topological insulators~\cite{dorin2021broadband} and nonreciprocal wave propagation~\cite{wu2018metastable,wu2019wave,adlakha2020frequency}. 

Recently, phase-modulated metasurfaces~\cite{assouar2018acoustic} have gained increasing research interest due to their ability to control low-frequency waves with compact and lightweight structures. A metasurface is a thin layer in the host medium composed of an array of subwavelength-scaled features, which can introduce an abrupt phase shift in the wave propagation path and tailor the wavefront based on generalized Snell’s law. The first studies on metasurface designs rely on linear properties of the unit structures to modulate the acoustic/elastic wavefront~\cite{PhysRevApplied.2.064002,cummer2016controlling,PhysRevLett.117.034302,zhu2020bifunctional,lin2021elastic}. For instance, metasurface designs based on Helmholtz-resonator\cite{lan2017manipulation} or coiling-up space structure~\cite{PhysRevApplied.2.064002} were proposed to control the acoustic wavefront of reflected and transmitted waves. To enhance wave control opportunities beyond conventional linear metasurfaces, researchers has recently introduced nonlinearity in the acoustic metasurface designs and achieved extraordinary and unconventional wave controllability. For example, Popa and Cummer~\cite{popa2014non} presented a single layer structure with Helmholtz-resonators and nonlinear electronic circuits to achieve  second-harmonic generation (SHG) for nonreciprocal wave propagation. Guo et al.~\cite{guo2018manipulating,guo2019frequency} proposed nonlinear acoustic metasurfaces based on mass-spring and rotating-square structures, and explored SHG in the reflected region, revealing that the local resonance effects can amplify SHG. 
%They succeeded in amplifying the SHG by setting the higher-order natural frequencies of the structure to be twice as high as the lower one.

While promising, the current nonlinear acoustic metasurface studies mainly focus on generating and maximizing the higher-order harmonics. Yet, the phase modulation and wavefront control capability of these structures have not been explored in the acoustic regime to the authors' knowledge. In addition, the resulting scattered wavefields often comprise a mixture of different harmonic components, which may weaken the efficiency of controlling them individually.

In this paper,  we advance the state of the art by introducing a novel nonlinear acoustic metasurface for simultaneous higher-harmonic generation and unconventional wavefront manipulation. We examine the amplitude-dependent behavior of the SHG, and the phase modulation capability of the proposed metasurface, proving that it can effectively generate second-harmonic wave and simultaneously modulate its wavefront for various functions, including wave deflecting, wave focusing and self-bending propagation, without influencing the propagation of fundamental wave components. In other words, we are the first to demultiplex the SHG from the fundamental wave component for individual acoustic wavefront control.

\begin{figure*}
\includegraphics[width=16cm]{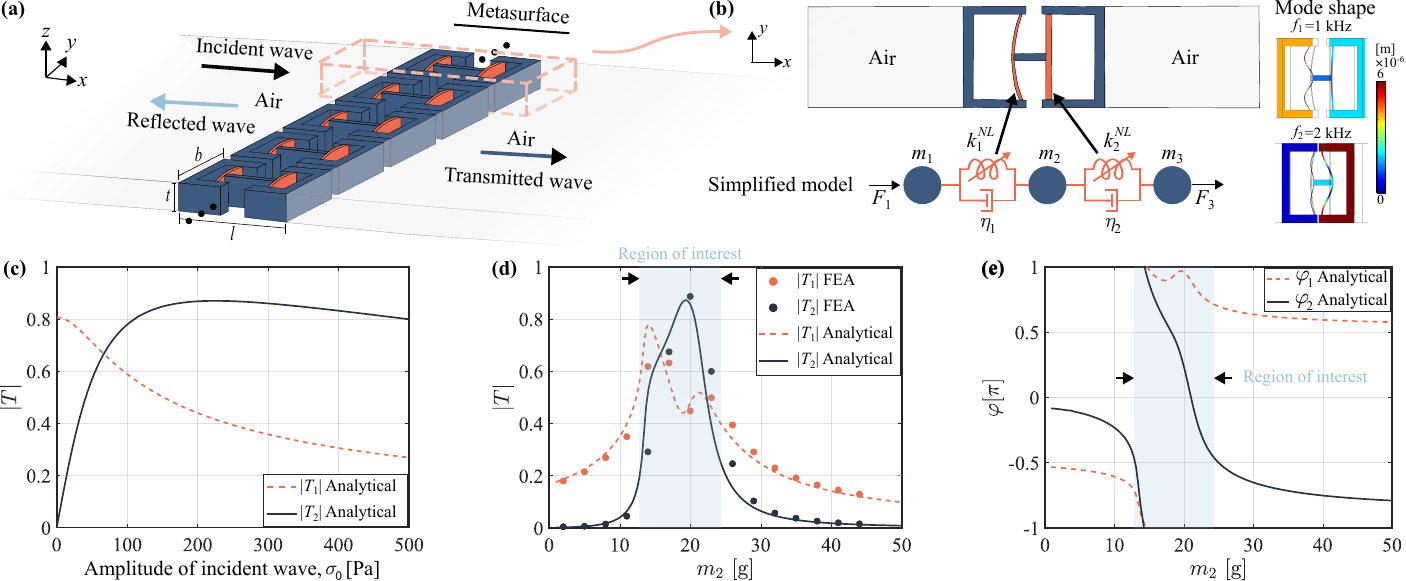}% Here is how to import EPS art
\caption{\label{fig:acoustic_waveguide} Schematic of the (\textbf{a}) nonlinear acoustic metasurface and (\textbf{b}) the one-dimensional waveguide. The inset in (\textbf{b}) depicts the mode shapes of the unit cell when $f_2=2f_1=2$ kHz. (\textbf{c}) The transmission ratios of both fundamental component ($|T_1|$) and SHG ($|T_2|$) under different amplitudes of the incident wave. (\textbf{d}) The transmission ratios and (\textbf{e}) phase modulation curve of fundamental wave component ($\varphi_1$) and SHG ($\varphi_2$) vary with middle mass value $m_2$.}
\end{figure*}

 The proposed nonlinear acoustic metasurface concept is depicted in Fig.~\ref{fig:acoustic_waveguide}(\textbf{a}), which is composed of an array of sub-wavelength unit cells, consisting of three lumped masses coupled via two curved beams. The unit cell has a length of $l = 50.8$ mm, a width of $b =50.8$ mm and a thickness of $t = 100$ mm.
We design the metasurface to control the acoustic wave propagating in air with mass density $\rho = 1.29$~kg/m$^3$ and sound speed $c=340$ m/s. We focus on the problem of acoustic wave transmission when a plane wave at a single frequency normally impinges on the left lumped mass, $m_1$, and propagates through the metasurface. To analytically formulate the problem, we establish a simplified mass-spring-damper system, where the mass of the metasurface unit is assumed to be concentrated at the three lumped masses $m_1, m_2$, and $m_3$, and each curved beam is modeled as an ideal nonlinear spring with damping, as shown in Fig.~\ref{fig:acoustic_waveguide}(\textbf{b}). 

Without loss of generality, the restoring forces of the nonlinear springs are assumed of the form, $F^{NL}_{i} (\Delta l)=k_{i}\Delta l+\alpha_{i} k_{i} \Delta l^2+\beta_{i} k_{i}\Delta l^3, i=1,2$, where $\Delta l$ is the deformation of the nonlinear springs, and $k_{i},~\alpha_{i}$,~and~$\beta_{i}$ are the stiffness parameters of the nonlinear springs.~These stiffness parameters are determined by fitting the force-displacement curves of the two curved beams obtained via the quasi-static analysis in COMSOL Multiphysics, yielding: $k_{1}=2.78\times10^5~\text{N/m}, \alpha_{1}=2.00\times10^3~\text{m}^{-1}, \beta_{1}=2.30\times10^6~\text{m}^{-2}, k_{2}=6.99\times10^5~\text{N/m}, \alpha_{2}=-510~\text{m}^{-1}, \beta_{2}=1.02\times10^6 \text{m}^{-2}$.
%$(k_{1}, \alpha_{1}, \beta_{1})=(2.78\times10^5$ N/m, $2.00\times10^3 \text{m}^{-1}, 2.30\times10^6 \text{m}^{-2})$, $(k_{2}, \alpha_{2}, \beta_{2})=(6.99\times10^5$ N/m, $-510 \text{m}^{-1}, 1.02\times10^6 \text{m}^{-2})$. 
We select the mass values $m_1=9.4~\text{g}, m_2=19.0~\text{g}, m_3=6.0~\text{g}$, resulting in a system with two non-zero natural frequencies satisfying $f_2 = 2f_1 = 2$ kHz. It is pointed out in previous studies \cite{guo2018manipulating,guo2019frequency} that under such doubling resonance frequency condition, i.e., $f_2 = 2f_1$, the SHG can be maximized when the system is excited at $f_1$. 

We assume a longitudinal wave with plane stress-wave field $\sigma_{inc}(x, t) = \sigma_0 \sin(2\pi f(t-\frac{x}{c}))$ normally incident from the left-hand side of the metasurface unit along the positive $x$-direction, where $\sigma_0$ is the amplitude of the incident wave and the incident frequency is set to be $f = f_1 = 1$ kHz. Accordingly, the reflected and transmitted stress-wave fields can be written as follows.
\begin{equation}
\label{eqn:ref and tran}
\sigma_{ref} = \sigma_{inc} +\rho c \frac{\partial u}{\partial t} ,  \sigma_{tr} = -\rho c \frac{\partial u}{\partial t} ,
\end{equation}
where $u(x,t)$ is the resulting displacement field in the air. $\sigma_{ref} $ and $\sigma_{tr} $ represent the reflected and transmitted stress-wave fields, respectively. Given that the metasurface is located at $x=0$ and its length $l$ is much less than the considered wavelength ($\lambda=340$ mm), the effective forces applied at $m_1$ and $m_3$ can then be determined by matching the boundary conditions on the metasurface at $x=0$, i.e., $F_1 = -(\sigma_{inc}|_{x=0 }+ \sigma_{ref}|_{x=0})S$, and $F_3 = \sigma_{tr}|_{x=0} S$, where $S = bt$ is the area of the flat surfaces contacting air. As such, the governing equations of the metasurface unit can be expressed as:
 %= -\big(2\sigma_0 \sin(2\pi f t)+\rho c \frac{du_1}{dt}\big)S =-\rho c \frac{du_3}{dt} S
\vspace{0pt}
\begin{subequations}
\label{GoverningEqn}
    \begin{equation}
    \setlength{\belowdisplayskip}{0pt} 
 \setlength{\abovedisplayskip}{0pt} 
    \begin{aligned}
        m_1 \frac{d^2 u_1}{dt^2} \ = &\  -\big(2\sigma_0 \sin(2\pi f t)  +\rho c \frac{du_1}{dt}\big)S - \\
        &F^{NL}_1 (u_1 - u_2) - \eta_1 \frac{d(u_1 - u_2)}{dt},
        \end{aligned}\\ 
        \end{equation}
        \begin{equation}
 \setlength{\belowdisplayskip}{0pt} 
 \setlength{\abovedisplayskip}{0pt} \setlength{\abovedisplayshortskip}{0pt}
    \begin{aligned}
        m_2 \frac{d^2 u_2}{dt^2} \ = &\ F^{NL}_1 (u_1-u_2)  + \eta_1 \frac{d(u_1 - u_2)}{dt} - \\
        &F^{NL}_2 (u_2-u_3)  -  \eta_2 \frac{d(u_2 - u_3)}{dt},
        \end{aligned}\\ 
            \end{equation}
             \begin{equation}
  \setlength{\belowdisplayskip}{0pt}  
   \setlength{\abovedisplayskip}{0pt}  
   \setlength{\abovedisplayshortskip}{0pt}
        \begin{aligned}
        m_3 \frac{d^2 u_3}{dt^2} \ =  & \ -\rho c S \frac{du_3}{dt} + F^{NL}_2(u_2 -u_3)+ \quad \\ & \eta_2 \frac{d(u_2-u_3)}{dt},
        \end{aligned}
       \end{equation}
\end{subequations}
where $u_1$, $u_2$ and $u_3$ are the displacements of $m_1$,~$m_2$~and~$m_3$, respectively. $\eta_1, \eta_2$ are the damping coefficients of the two nonlinear springs. Then, Eq.~\ref{GoverningEqn} is solved analytically by adopting the Harmonic Balance Method (HBM) with the solutions $u_i \ (i= 1, 2, 3)$ assumed as:
%in the form of a sum of all harmonics. That is, we have the following solutions.
\begin{equation}
    \label{Solutions}
    u_i(t) = u_{i}^0 + \sum_{n=1}^{N} \big[ C_{i}^n \cos{(2\pi n f t)} + S_{i}^n \sin{(2\pi n f t)}\big],
\end{equation}
where $u_{i}^0$ represents the constant terms, $C_{i}^n$ and $S_{i}^n$ are the magnitudes of the sinusoidal terms, cosine and sine, respectively, and $N$ is the number of harmonics truncated. By substituting Eq.~\ref{Solutions} into Eq.~\ref{GoverningEqn} and matching the coefficients for different harmonics, we obtain a set of nonlinear equations with respect to the unknown coefficients $u_{i}^0$, $C_{i}^n$ and $S_{i}^n$, which can be numerically solved by Newton-Raphson method~\cite{ypma1995historical}. Then, the complex transmission coefficients of each harmonic component can be obtained by substituting Eq.~\ref{Solutions} into Eq.~\ref{eqn:ref and tran}, yielding:
\begin{equation}
    \label{TranCoe}
    T_n = \frac{2\pi n f \rho c }{\sigma_0} (C_{i}^n-j S_{i}^n), \ n = 1,2,..., N,
\end{equation}
where $j$ is the imaginary unit. The transmission ratio, $|T_n|$, of the $n^{th}$ order harmonics (i.e., the ratio between the amplitude of the transmitted $n^{th}$ order harmonic wave and incident wave), and its corresponding phase shift, $\varphi_n$, can be calculated by taking the magnitude and angle of $T_n$, respectively.

\begin{figure*}
\includegraphics[width=18cm]{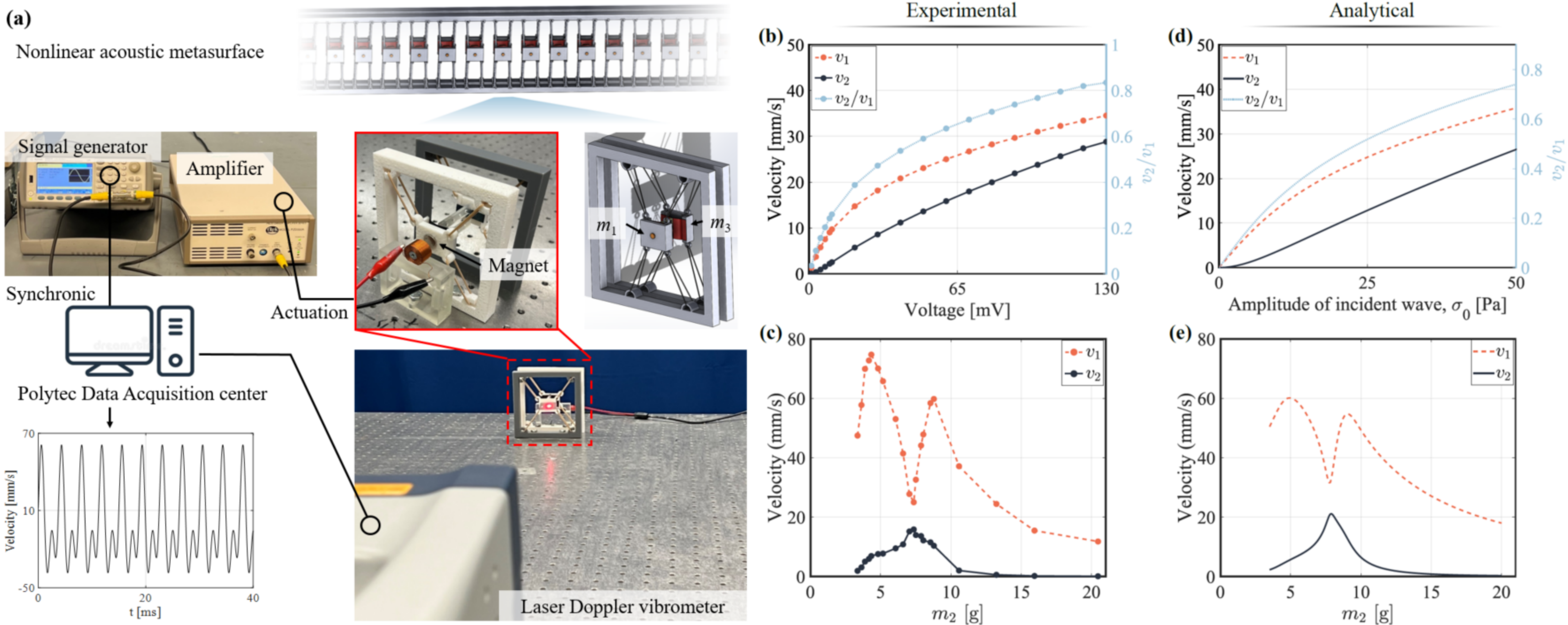}% Here is how to import EPS art
\caption{\label{fig:exp} (\textbf{a}) The nonlinear acoustic metasurface and the waveguide experimental configuration. (\textbf{b})-(\textbf{e}) Comparison between the experimental and analytical velocity responses under harmonic excitation at $f=f_1=266.4$ Hz. (\textbf{b}) Measured velocity of right mass block, $m_3$, under different voltage inputs, indicating stronger nonlinearity with larger excitation. (\textbf{c}) Measured velocity responses under different middle mass value, $m_2$, and fixed voltage 60 mV. (\textbf{d}) Analytical prediction of the amplitude-dependent response with experimentally measured curved beam parameters. (\textbf{e}) Analytical velocity responses with incident amplitude $\sigma_0=40$ Pa and damping coefficients $\eta_1 = \eta_2 = 0.15$ under different $m_2$, demonstrating a similar trend to the waveguide transmission modulation experiments, as shown in (\textbf{c}).}
\end{figure*}

%\subsection{Results and validation}
%\label{sec:Results and validation}

With this analytical model, we investigate the transmission characteristics, i.e., the transmission ratio and phase shift, of the undamped acoustic waveguide, as depicted in Figs.~\ref{fig:acoustic_waveguide}(\textbf{c})-(\textbf{e}). We first explore the amplitude-dependent behavior of the acoustic waveguide under different incident amplitudes, $\sigma_0$. As shown in Fig.~\ref{fig:acoustic_waveguide}(\textbf{c}), the fundamental wave component ($f_1$) dominates the transmitted wavefield when $\sigma_0 \approx 0$ Pa, indicating a nearly linear response.~As $\sigma_0$ increases, the nonlinear effect occurs, leading to gradually enhanced SHG. The transmission ratio of SHG tends to saturate and reaches a maximal value of 0.84 when $\sigma_0 \approx 200$ Pa. Since all the other higher-harmonic components ($n\geq3$) are negligible, most of the acoustic energy is converted to the SHG, demonstrating that the proposed metasurface unit cell provides an effective and efficient platform for SHG. The mechanism for the significant SHG in the transmitted region can be further interpreted by examining the mode shapes of the acoustic waveguide, as shown in the inset of Fig.~\ref{fig:acoustic_waveguide}(\textbf{b}). It is observed that the right mass $m_3$ has a large motion in the second resonant mode, i.e., $f_2$, while the left mass $m_1$ remains almost steady, indicating that the second-harmonic wave tends to propagate into the transmitted region instead of being reflected.

Moreover, we explore the waveguide's capability to modulate the wave transmission characteristics by tuning the parameters of the unit cell, specifically, the middle mass value $m_2$. We vary $m_2$ from 1 g to 50 g and fix other parameters with $\sigma_0 = 200$ Pa for higher SHG. The transmission ratios of fundamental and second-harmonic components change dramatically with different $m_2$ due to the shift in natural resonances of the metasurface unit, as shown in Fig.~\ref{fig:acoustic_waveguide}(\textbf{d}). As expected, the maximum SHG occurs when $m_2=19$ g, i.e., when $f_2 = 2f_1$. Apart from the transmission ratios, the phase modulation capability is another key factor in metasurface design. The phase gradient should span the range from $-\pi$ to $\pi$ in order to effectively control and manipulate the acoustic field. The phase modulation results in Fig.~\ref{fig:acoustic_waveguide}(\textbf{e}) show that our design can introduce different phase shift profiles for the fundamental and second-harmonic waves by tailoring $m_2$, which indicates the feasibility to achieve diverse wavefront control over different harmonic components. It is observed that the phase shift curve of the SHG covers the entire 2$\pi$ phase range, revealing that any type of anomalous wavefront control over the SHG can be realized, while the phase shift of the fundamental component is insensitive to the change of $m_2$ in the range of interest. Note that this interesting phenomenon is utilized in the following metasurface designs to split the SHG from the fundamental component for individual control. Furthermore, through finite element analysis (FEA) in COMSOL Multiphysics, the transmission ratios of both harmonics are obtained as plotted in Fig.~\ref{fig:acoustic_waveguide}(\textbf{d}), demonstrating an excellent agreement with the analytical solutions.

Next, we fabricate and experimentally test the acoustic waveguide to validate the analytical and numerical predictions of its transmission characteristics. The mass blocks are fabricated through Form 3 SLA 3D Printer and connected via curved beams made of 1095 spring steels. The stiffness parameters of the curved beams are tested through Tensile Testing Machines, yielding: $k_{1}=2.60\times10^4~\text{N/m},~\alpha_{1}=3.84\times10^3~\text{m}^{-1},~\beta_{1}=8.53\times10^5~\text{m}^{-2},~k_{2}=3.28\times10^4~\text{N/m},~\alpha_{2}=141~\text{m}^{-1},~\beta_{2}=1.03\times10^3~\text{m}^{-2}$.
%$(k_{1}, \alpha_{1}, \beta_{1})=(2.60\times 10^4 \text{N/m},3.84\times^3  \text{m}^{-1},8.53 \times 10^5 \text{m}^{-2} )$, $(k_{2}, \alpha_{2}, \beta_{2})=(3.28 \times 10^4 \text{N/m},141 \text{m}^{-1} ,1.03 \times 10^3 \text{m}^{-2})$. 
Accordingly, the mass values $m_1=14.42~\text{g},~m_2=7.35~\text{g},~m_3=7.65~\text{g}$ are selected to guarantee the second natural frequency doubles the fundamental one, i.e., $f_1=f_2/2=266.4$ Hz. The waveguide experiment is conducted with the unit cell hung with rubber bands, as depicted in Fig.~\ref{fig:exp}(\textbf{a}). A circular magnet with 12.7 mm diameter and 1.5 mm thickness is bonded at the center of $m_1$.  The excitation signal is generated by the waveform generator (Agilent 33522A) and sent to an electromagnet copper magnetic coil through an amplifier (TREK PZD350A) to apply tunable harmonic forces to $m_1$. We measure the out-of-plane velocity of $m_3$ via Polytec PSV-500 scanning laser Doppler vibrometer (LDV) to obtain the transmitted waves, as demonstrated in the inset of Fig.~\ref{fig:exp}(\textbf{a}), and apply the Fast Fourier transformation technique to obtain the amplitudes of fundamental and second-harmonic components.
 
To characterize the amplitude-dependent behavior of our experimental platform, we tune the voltage in the signal generator and record the transmitted wave under the same excitation frequency $f = f_1 = 266.4$ Hz, as plotted in Fig.~\ref{fig:exp} (\textbf{b}). The results verify that our system can realize significant SHG in the transmitted region, and higher input voltage leads to a larger ratio between the velocity amplitudes of SHG and the fundamental component ($v_2/v_1$), indicating that more energy is converted into SHG as the system nonlinearity increases. In addition, we experimentally measure the performance of our system under different middle mass values with the same excitation, as presented in Fig.~\ref{fig:exp} (\textbf{c}). The SHG is maximized when $m_2=7.35$ g, which corresponds to the arrangement with $f = f_1=f_2/2=266.4$ Hz. Meanwhile, a local drop is observed in the fundamental wave velocity around the peak of SHG since more energy is transformed into SHG around this region.  In addition, Figs.~\ref{fig:exp} (\textbf{d}) and (\textbf{e}) show the corresponding analytical results with the experimentally measured structural parameters and proper selection of incident amplitude and damping coefficients. Our experimental results and the analytical solutions match each other well in terms of the amplitude-dependent behavior and the modulation of the transmission characteristics, revealing that our experimental platform can be utilized to provide an excellent physical implementation of the proposed nonlinear acoustic waveguide for SHG and transmitted wave manipulation as analyzed and designed by the theoretical framework.

\begin{figure}
\includegraphics[width=6cm]{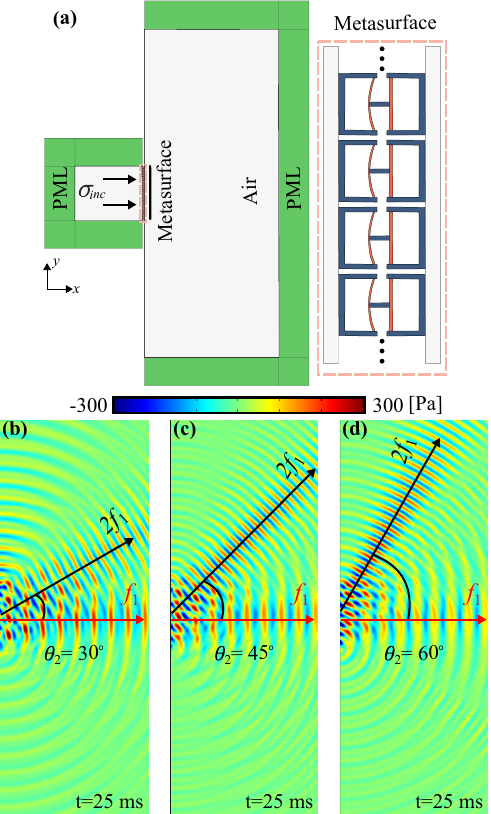}% Here is how to import EPS art
\caption{\label{fig:deflecting} (\textbf{a}) Schematic of the nonlinear acoustic metasurface placed in the air. (\textbf{b})-(\textbf{d})  The transmitted acoustic pressure fields of deflecting metasurfaces with desired angle (\textbf{b}) $\theta_2=30^\circ$, (\textbf{c}) $\theta_2=45^\circ$, and (\textbf{d}) $\theta_2=60^\circ$.}
\end{figure}

%\section{Simulation}
Having the analytical model of the nonlinear waveguide validated through both numerical simulations and experiments, we design acoustic metasurfaces to manipulate the wavefront of the transmitted wavefield based on the phase shift profiles shown in Fig.~\ref{fig:acoustic_waveguide}(\textbf{e}). The designed metasurface consists of an array of 25 units with a gap of 2 mm. By tailoring the middle mass value $m_2$ along the metasurface, we present different metasurfaces to control the normally incident acoustic wave at 1 kHz and achieve full wavefront control of the SHG. In all metasurface designs, the fundamental wave components remain in the same propagating direction as the incident waves since most of the metasurface units introduce a constant phase shift $\varphi_1(y) \approx \pi$, where $y$ represents the vertical coordinates with the origin located at the center of the metasurface, while the SHG will be split from the fundamental component for various types of anomalous wavefront control. The performance of the proposed metasurfaces is validated via finite element simulations conducted in COMSOL Multiphysics, where the metasurfaces are surrounded by the perfectly matched layers (PML) to avoid the reflection from the boundary and simulate the infinite acoustic domain, as depicted in Fig.~\ref{fig:deflecting}(\textbf{a}). The first type of nonlinear metasurface is designed to demultiplex different frequency components in the transmitted region by guiding the generated second-harmonic wave in different desired angles, $\theta_2$, according to the design principle based on the generalized Snell's law: $\varphi_2(y)=-y k_2 \sin{\theta_2} $, where $k_2$ is the wavenumber of the second-harmonic component. To this end, we design three metasurfaces with theoretical deflecting angles of $\theta_2=30^\circ$, $\theta_2=45^\circ$, and $\theta_2=60^\circ$ for the second-harmonic wave. Figures~\ref{fig:deflecting}(\textbf{b})-(\textbf{d}) show the resulting scattered acoustic pressure fields in the transmitted region, where significant SHG is realized and the incident waves are successfully split into two beams as designed. One beam maintains the original propagating direction with the same wavelength as the incident wave (i.e., the fundamental wave component). The other is deflected by the desired angle, $\theta_2$, and propagates with half of the original wavelength (i.e., the second-harmonic component). Therefore, the proposed metasurfaces can simultaneously achieve significant SHG and frequency demultiplexing.

\begin{figure}
\includegraphics[width=8.5cm]{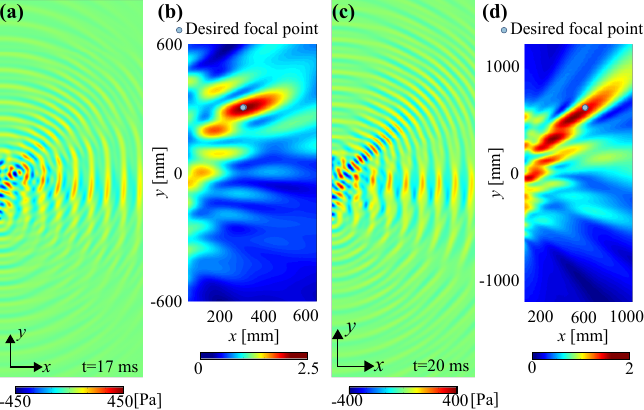}% Here is how to import EPS art
\caption{\label{fig:focusing} Focusing metasurface designs with different desired focal points. (\textbf{a}) The transmitted acoustic pressure field and (\textbf{b}) normalized RMS pressure field of the focusing metasurface with desired focal point $(\it{f_x}, \it{f_y})$=(304.8,304.8) mm, demonstrating amplification effect of 2.12 times in the focal area. (\textbf{c}) The transmitted acoustic pressure field and (\textbf{d}) normalized RMS pressure field of the focusing metasurface with desired focal point $(\it{f_x}, \it{f_y})$=(609.6,609.6) mm, closer to the upper boundary of the metasurface.}
\end{figure}

 Asides from deflecting metasurface, the focusing metasurface is of particular interest in engineering applications requiring localized high-intensity wave energy. To focus the SHG at $(\it{f_x}, \it{f_y})$ away from the center of the metasurface, a hyperbolic-type phase profile $\varphi_2(y)=k_2 (\sqrt{f_x^2+(y-f_y)^2}-\sqrt{f_x^2+f_y^2})$ is required to form a semicircle equiphase surface in the transmitted region \cite{zhu2015dispersionless}. Here we demonstrate a focusing metasurface design to localize the SHG at $(\it{f_x}, \it{f_y})$=(304.8, 304.8) mm. The transmitted acoustic pressure field is shown in Fig.~\ref{fig:focusing}(\textbf{a}), and the root-mean-square (RMS) acoustic pressure results are calculated and normalized by the background acoustic pressure as plotted in ~Fig.~\ref{fig:focusing}(\textbf{b}). The maximum acoustic pressure intensity occurs at (306, 300) mm with only (0.39$\%$, 1.57$\%$) deviation away from the desired focal point, and the acoustic pressure is amplified by 2.12 times in the focal area. Hence, the nonlinear focusing metasurface can focus the SHG at the desired region, offering great potential in imaging small objects for various applications (such as bio-medical or structural health monitoring). In addition, the performance of the metasurface is examined to focus the SHG at a further point, i.e., (609.6, 609.6) mm, as presented in Figs.~\ref{fig:focusing}(\textbf{c}) and (\textbf{d}), 
 where a larger focal region is observed with a smaller amplification factor of 1.64 at (506,~520) mm. The focusing performance at the further focal point decreases due to the limitation in the aperture size of the metasurface, which can be enhanced by increasing the number of the metasurface units.

\begin{figure}
\includegraphics[width=8.5cm]{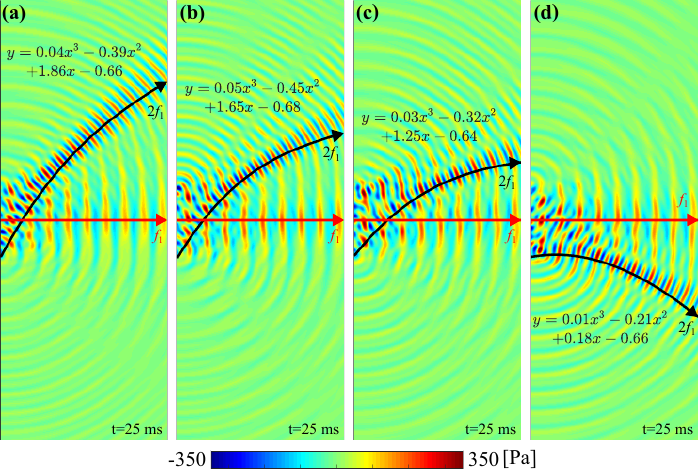}% Here is how to import EPS art
\caption{\label{fig:selfbending} The transmitted acoustic pressure fields of different self-bending beam propagation designs with desired trajectory (\textbf{a}) $y=0.04x^3-0.39x^2+1.86x-0.66$, (\textbf{b}) $y=0.05x^3-0.45x^2+1.65x-0.68$, (\textbf{c}) $y=0.03x^3-0.32x^2+1.25x-0.64$, and (\textbf{d}) $y=0.01x^3-0.21x^2+0.18x-0.66$.}
\end{figure}

 To further demonstrate the versatility of our proposed nonlinear designs, the metasurface is used to guide the generated second-harmonic wave to propagate along a self-bending trajectory. To form a self-bending beam propagation, all the rays in the transmitted region should be tangent to the desired caustic trajectory~\cite{froehly2011arbitrary}. Accordingly, the phase gradient profile of the metasurface can be determined via Legendre transform technique~\cite{zhang2014generation}. We design four metasurfaces to guide the SHG to propagate along different cubic caustic trajectories considering the aperture size. The resulting acoustic pressure fields are shown in Fig.~\ref{fig:selfbending}, where the SHG is successfully bent and propagates along the desired self-bending trajectory. Hence, the nonlinear acoustic metasurface does not only demultiplex different frequency components but also guide the SHG to propagate in any desired trajectory. Note that all the metasurface designs are achieved via tailoring middle mass value $m_2$, which can be realized by adding/subtracting additional mass units, enabling highly tunable nonlinear metasurface designs for broadband frequency wave control.

In summary, we propose and explore a novel nonlinear acoustic metasurface consisting of structural modules with curved beams to tailor and harness SHG for anomalous wavefront control. Our innovation and findings include: (i) a nonlinear waveguide mechanism for achieving significant SHG propagating into the transmission region, (ii) insights from amplitude-dependent analysis of transmission and phase modulation by tailoring the structural parameters of the proposed nonlinear unit cell, and (iii) design of nonlinear acoustic metasurfaces to demultiplex different harmonics and control the second-harmonic wavefront for diverse functions, including wave deflecting, wave focusing, and self-bending propagation. With the advanced theoretical, numerical, and experimental efforts, this study uncovers unconventional wave manipulation and control via nonlinear metasurfaces and creates new potentials toward a broad range of engineering applications, e.g., phonon computing~\cite{sklan2015splash} and ultrasonic diagnosis and therapy in biomedicine, where higher-order harmonic generation can be harnessed to image smaller objects~\cite{rudenko2006giant}.

%\begin{acknowledgments}
This work was supported by the National Science Foundation under Grant No. CMMI-1933436 and the Air Force Office of Scientific Research under Grant FA9550-21-1-0032.
%\end{acknowledgments}
\section*{AUTHOR DECLARATIONS}
\vspace{-10pt}
\subsection*{Author Contributions}
%\subsubsection*{\normalfont{\textbf{\textsf{Author contributions}}}}
\vspace{-15pt}
Z. L. and Y. Z. contributed equally to this work.
\vspace{-10pt}
\section*{Data Availability Statement}
\vspace{-10pt}
The data that support the findings of
this study are available from the
corresponding author upon reasonable
request.

\nocite{*}
\bibliography{aplacoustic}% Produces the bibliography via BibTeX.

\begin{thebibliography}{24}%
\makeatletter
\providecommand \@ifxundefined [1]{%
 \@ifx{#1\undefined}
}%
\providecommand \@ifnum [1]{%
 \ifnum #1\expandafter \@firstoftwo
 \else \expandafter \@secondoftwo
 \fi
}%
\providecommand \@ifx [1]{%
 \ifx #1\expandafter \@firstoftwo
 \else \expandafter \@secondoftwo
 \fi
}%
\providecommand \natexlab [1]{#1}%
\providecommand \enquote  [1]{``#1''}%
\providecommand \bibnamefont  [1]{#1}%
\providecommand \bibfnamefont [1]{#1}%
\providecommand \citenamefont [1]{#1}%
\providecommand \href@noop [0]{\@secondoftwo}%
\providecommand \href [0]{\begingroup \@sanitize@url \@href}%
\providecommand \@href[1]{\@@startlink{#1}\@@href}%
\providecommand \@@href[1]{\endgroup#1\@@endlink}%
\providecommand \@sanitize@url [0]{\catcode `\\12\catcode `\$12\catcode
  `\&12\catcode `\#12\catcode `\^12\catcode `\_12\catcode `\%12\relax}%
\providecommand \@@startlink[1]{}%
\providecommand \@@endlink[0]{}%
\providecommand \url  [0]{\begingroup\@sanitize@url \@url }%
\providecommand \@url [1]{\endgroup\@href {#1}{\urlprefix }}%
\providecommand \urlprefix  [0]{URL }%
\providecommand \Eprint [0]{\href }%
\providecommand \doibase [0]{http://dx.doi.org/}%
\providecommand \selectlanguage [0]{\@gobble}%
\providecommand \bibinfo  [0]{\@secondoftwo}%
\providecommand \bibfield  [0]{\@secondoftwo}%
\providecommand \translation [1]{[#1]}%
\providecommand \BibitemOpen [0]{}%
\providecommand \bibitemStop [0]{}%
\providecommand \bibitemNoStop [0]{.\EOS\space}%
\providecommand \EOS [0]{\spacefactor3000\relax}%
\providecommand \BibitemShut  [1]{\csname bibitem#1\endcsname}%
\let\auto@bib@innerbib\@empty
%</preamble>
\bibitem [{\citenamefont {Barnhart}\ \emph {et~al.}(2019)\citenamefont
  {Barnhart}, \citenamefont {Xu}, \citenamefont {Chen}, \citenamefont {Zhang},
  \citenamefont {Song},\ and\ \citenamefont
  {Huang}}]{barnhart2019experimental}%
  \BibitemOpen
  \bibfield  {author} {\bibinfo {author} {\bibfnamefont {M.~V.}\ \bibnamefont
  {Barnhart}}, \bibinfo {author} {\bibfnamefont {X.}~\bibnamefont {Xu}},
  \bibinfo {author} {\bibfnamefont {Y.}~\bibnamefont {Chen}}, \bibinfo {author}
  {\bibfnamefont {S.}~\bibnamefont {Zhang}}, \bibinfo {author} {\bibfnamefont
  {J.}~\bibnamefont {Song}}, \ and\ \bibinfo {author} {\bibfnamefont
  {G.}~\bibnamefont {Huang}},\ }\bibfield  {title} {\enquote {\bibinfo {title}
  {Experimental demonstration of a dissipative multi-resonator metamaterial for
  broadband elastic wave attenuation},}\ }\href@noop {} {\bibfield  {journal}
  {\bibinfo  {journal} {J. Sound Vib.}\ }\textbf {\bibinfo {volume} {438}},\
  \bibinfo {pages} {1--12} (\bibinfo {year} {2019})}\BibitemShut {NoStop}%
\bibitem [{\citenamefont {Lin}, \citenamefont {Al~Ba’ba’a},\ and\
  \citenamefont {Tol}(2021)}]{lin2021piezoelectric}%
  \BibitemOpen
  \bibfield  {author} {\bibinfo {author} {\bibfnamefont {Z.}~\bibnamefont
  {Lin}}, \bibinfo {author} {\bibfnamefont {H.}~\bibnamefont {Al~Ba’ba’a}},
  \ and\ \bibinfo {author} {\bibfnamefont {S.}~\bibnamefont {Tol}},\ }\bibfield
   {title} {\enquote {\bibinfo {title} {Piezoelectric metastructures for
  simultaneous broadband energy harvesting and vibration suppression of
  traveling waves},}\ }\href@noop {} {\bibfield  {journal} {\bibinfo  {journal}
  {Smart Mater. Struct.}\ }\textbf {\bibinfo {volume} {30}},\ \bibinfo {pages}
  {075037} (\bibinfo {year} {2021})}\BibitemShut {NoStop}%
\bibitem [{\citenamefont {Tol}, \citenamefont {Degertekin},\ and\ \citenamefont
  {Erturk}(2017)}]{tol2017phononic}%
  \BibitemOpen
  \bibfield  {author} {\bibinfo {author} {\bibfnamefont {S.}~\bibnamefont
  {Tol}}, \bibinfo {author} {\bibfnamefont {F.~L.}\ \bibnamefont {Degertekin}},
  \ and\ \bibinfo {author} {\bibfnamefont {A.}~\bibnamefont {Erturk}},\
  }\bibfield  {title} {\enquote {\bibinfo {title} {Phononic crystal luneburg
  lens for omnidirectional elastic wave focusing and energy harvesting},}\
  }\href@noop {} {\bibfield  {journal} {\bibinfo  {journal} {Appl. Phys.
  Lett.}\ }\textbf {\bibinfo {volume} {111}},\ \bibinfo {pages} {013503}
  (\bibinfo {year} {2017})}\BibitemShut {NoStop}%
\bibitem [{\citenamefont {Hu}\ \emph {et~al.}(2021)\citenamefont {Hu},
  \citenamefont {Tang}, \citenamefont {Liang}, \citenamefont {Lan},\ and\
  \citenamefont {Das}}]{hu2021acoustic}%
  \BibitemOpen
  \bibfield  {author} {\bibinfo {author} {\bibfnamefont {G.}~\bibnamefont
  {Hu}}, \bibinfo {author} {\bibfnamefont {L.}~\bibnamefont {Tang}}, \bibinfo
  {author} {\bibfnamefont {J.}~\bibnamefont {Liang}}, \bibinfo {author}
  {\bibfnamefont {C.}~\bibnamefont {Lan}}, \ and\ \bibinfo {author}
  {\bibfnamefont {R.}~\bibnamefont {Das}},\ }\bibfield  {title} {\enquote
  {\bibinfo {title} {Acoustic-elastic metamaterials and phononic crystals for
  energy harvesting: A review},}\ }\href@noop {} {\bibfield  {journal}
  {\bibinfo  {journal} {Smart Mater. Struct.}\ } (\bibinfo {year}
  {2021})}\BibitemShut {NoStop}%
\bibitem [{\citenamefont {Dorin}\ and\ \citenamefont
  {Wang}(2021)}]{dorin2021broadband}%
  \BibitemOpen
  \bibfield  {author} {\bibinfo {author} {\bibfnamefont {P.}~\bibnamefont
  {Dorin}}\ and\ \bibinfo {author} {\bibfnamefont {K.~W.}\ \bibnamefont
  {Wang}},\ }\bibfield  {title} {\enquote {\bibinfo {title} {Broadband
  frequency and spatial on-demand tailoring of topological wave propagation
  harnessing piezoelectric metamaterials},}\ }\href@noop {} {\bibfield
  {journal} {\bibinfo  {journal} {Front. Mater.}\ ,\ \bibinfo {pages} {409}}
  (\bibinfo {year} {2021})}\BibitemShut {NoStop}%
\bibitem [{\citenamefont {Wu}, \citenamefont {Zheng},\ and\ \citenamefont
  {Wang}(2018)}]{wu2018metastable}%
  \BibitemOpen
  \bibfield  {author} {\bibinfo {author} {\bibfnamefont {Z.}~\bibnamefont
  {Wu}}, \bibinfo {author} {\bibfnamefont {Y.}~\bibnamefont {Zheng}}, \ and\
  \bibinfo {author} {\bibfnamefont {K.~W.}\ \bibnamefont {Wang}},\ }\bibfield
  {title} {\enquote {\bibinfo {title} {Metastable modular metastructures for
  on-demand reconfiguration of band structures and nonreciprocal wave
  propagation},}\ }\href@noop {} {\bibfield  {journal} {\bibinfo  {journal}
  {Phys. Rev. E}\ }\textbf {\bibinfo {volume} {97}},\ \bibinfo {pages} {022209}
  (\bibinfo {year} {2018})}\BibitemShut {NoStop}%
\bibitem [{\citenamefont {Wu}\ and\ \citenamefont {Wang}(2019)}]{wu2019wave}%
  \BibitemOpen
  \bibfield  {author} {\bibinfo {author} {\bibfnamefont {Z.}~\bibnamefont
  {Wu}}\ and\ \bibinfo {author} {\bibfnamefont {K.~W.}\ \bibnamefont {Wang}},\
  }\bibfield  {title} {\enquote {\bibinfo {title} {On the wave propagation
  analysis and supratransmission prediction of a metastable modular
  metastructure for non-reciprocal energy transmission},}\ }\href@noop {}
  {\bibfield  {journal} {\bibinfo  {journal} {J. Sound Vib.}\ }\textbf
  {\bibinfo {volume} {458}},\ \bibinfo {pages} {389--406} (\bibinfo {year}
  {2019})}\BibitemShut {NoStop}%
\bibitem [{\citenamefont {Adlakha}\ \emph {et~al.}(2020)\citenamefont
  {Adlakha}, \citenamefont {Moghaddaszadeh}, \citenamefont {Attarzadeh},
  \citenamefont {Aref},\ and\ \citenamefont {Nouh}}]{adlakha2020frequency}%
  \BibitemOpen
  \bibfield  {author} {\bibinfo {author} {\bibfnamefont {R.}~\bibnamefont
  {Adlakha}}, \bibinfo {author} {\bibfnamefont {M.}~\bibnamefont
  {Moghaddaszadeh}}, \bibinfo {author} {\bibfnamefont {M.~A.}\ \bibnamefont
  {Attarzadeh}}, \bibinfo {author} {\bibfnamefont {A.}~\bibnamefont {Aref}}, \
  and\ \bibinfo {author} {\bibfnamefont {M.}~\bibnamefont {Nouh}},\ }\bibfield
  {title} {\enquote {\bibinfo {title} {Frequency selective wave beaming in
  nonreciprocal acoustic phased arrays},}\ }\href@noop {} {\bibfield  {journal}
  {\bibinfo  {journal} {Sci. Rep.}\ }\textbf {\bibinfo {volume} {10}},\
  \bibinfo {pages} {1--14} (\bibinfo {year} {2020})}\BibitemShut {NoStop}%
\bibitem [{\citenamefont {Assouar}\ \emph {et~al.}(2018)\citenamefont
  {Assouar}, \citenamefont {Liang}, \citenamefont {Wu}, \citenamefont {Li},
  \citenamefont {Cheng},\ and\ \citenamefont {Jing}}]{assouar2018acoustic}%
  \BibitemOpen
  \bibfield  {author} {\bibinfo {author} {\bibfnamefont {B.}~\bibnamefont
  {Assouar}}, \bibinfo {author} {\bibfnamefont {B.}~\bibnamefont {Liang}},
  \bibinfo {author} {\bibfnamefont {Y.}~\bibnamefont {Wu}}, \bibinfo {author}
  {\bibfnamefont {Y.}~\bibnamefont {Li}}, \bibinfo {author} {\bibfnamefont
  {J.~C.}\ \bibnamefont {Cheng}}, \ and\ \bibinfo {author} {\bibfnamefont
  {Y.}~\bibnamefont {Jing}},\ }\bibfield  {title} {\enquote {\bibinfo {title}
  {Acoustic metasurfaces},}\ }\href@noop {} {\bibfield  {journal} {\bibinfo
  {journal} {Nat. Rev. Mater.}\ }\textbf {\bibinfo {volume} {3}},\ \bibinfo
  {pages} {460--472} (\bibinfo {year} {2018})}\BibitemShut {NoStop}%
\bibitem [{\citenamefont {Li}\ \emph {et~al.}(2014)\citenamefont {Li},
  \citenamefont {Jiang}, \citenamefont {Li}, \citenamefont {Liang},
  \citenamefont {Zou}, \citenamefont {Yin},\ and\ \citenamefont
  {Cheng}}]{PhysRevApplied.2.064002}%
  \BibitemOpen
  \bibfield  {author} {\bibinfo {author} {\bibfnamefont {Y.}~\bibnamefont
  {Li}}, \bibinfo {author} {\bibfnamefont {X.}~\bibnamefont {Jiang}}, \bibinfo
  {author} {\bibfnamefont {R.~Q.}\ \bibnamefont {Li}}, \bibinfo {author}
  {\bibfnamefont {B.}~\bibnamefont {Liang}}, \bibinfo {author} {\bibfnamefont
  {X.~Y.}\ \bibnamefont {Zou}}, \bibinfo {author} {\bibfnamefont {L.~L.}\
  \bibnamefont {Yin}}, \ and\ \bibinfo {author} {\bibfnamefont {J.~C.}\
  \bibnamefont {Cheng}},\ }\bibfield  {title} {\enquote {\bibinfo {title}
  {Experimental realization of full control of reflected waves with
  subwavelength acoustic metasurfaces},}\ }\href {\doibase
  10.1103/PhysRevApplied.2.064002} {\bibfield  {journal} {\bibinfo  {journal}
  {Phys. Rev. Appl.}\ }\textbf {\bibinfo {volume} {2}},\ \bibinfo {pages}
  {064002} (\bibinfo {year} {2014})}\BibitemShut {NoStop}%
\bibitem [{\citenamefont {Cummer}, \citenamefont {Christensen},\ and\
  \citenamefont {Al{\`u}}(2016)}]{cummer2016controlling}%
  \BibitemOpen
  \bibfield  {author} {\bibinfo {author} {\bibfnamefont {S.~A.}\ \bibnamefont
  {Cummer}}, \bibinfo {author} {\bibfnamefont {J.}~\bibnamefont {Christensen}},
  \ and\ \bibinfo {author} {\bibfnamefont {A.}~\bibnamefont {Al{\`u}}},\
  }\bibfield  {title} {\enquote {\bibinfo {title} {Controlling sound with
  acoustic metamaterials},}\ }\href@noop {} {\bibfield  {journal} {\bibinfo
  {journal} {Nat. Rev. Mater.}\ }\textbf {\bibinfo {volume} {1}},\ \bibinfo
  {pages} {1--13} (\bibinfo {year} {2016})}\BibitemShut {NoStop}%
\bibitem [{\citenamefont {Zhu}\ and\ \citenamefont
  {Semperlotti}(2016)}]{PhysRevLett.117.034302}%
  \BibitemOpen
  \bibfield  {author} {\bibinfo {author} {\bibfnamefont {H.}~\bibnamefont
  {Zhu}}\ and\ \bibinfo {author} {\bibfnamefont {F.}~\bibnamefont
  {Semperlotti}},\ }\bibfield  {title} {\enquote {\bibinfo {title} {Anomalous
  refraction of acoustic guided waves in solids with geometrically tapered
  metasurfaces},}\ }\href {\doibase 10.1103/PhysRevLett.117.034302} {\bibfield
  {journal} {\bibinfo  {journal} {Phys. Rev. Lett.}\ }\textbf {\bibinfo
  {volume} {117}},\ \bibinfo {pages} {034302} (\bibinfo {year}
  {2016})}\BibitemShut {NoStop}%
\bibitem [{\citenamefont {Zhu}\ \emph {et~al.}(2020)\citenamefont {Zhu},
  \citenamefont {Cao}, \citenamefont {Merkel}, \citenamefont {Fan},\ and\
  \citenamefont {Assouar}}]{zhu2020bifunctional}%
  \BibitemOpen
  \bibfield  {author} {\bibinfo {author} {\bibfnamefont {Y.}~\bibnamefont
  {Zhu}}, \bibinfo {author} {\bibfnamefont {L.}~\bibnamefont {Cao}}, \bibinfo
  {author} {\bibfnamefont {A.}~\bibnamefont {Merkel}}, \bibinfo {author}
  {\bibfnamefont {S.-W.}\ \bibnamefont {Fan}}, \ and\ \bibinfo {author}
  {\bibfnamefont {B.}~\bibnamefont {Assouar}},\ }\bibfield  {title} {\enquote
  {\bibinfo {title} {Bifunctional superlens for simultaneous flexural and
  acoustic wave superfocusing},}\ }\href@noop {} {\bibfield  {journal}
  {\bibinfo  {journal} {Appl. Phys. Lett.}\ }\textbf {\bibinfo {volume}
  {116}},\ \bibinfo {pages} {253502} (\bibinfo {year} {2020})}\BibitemShut
  {NoStop}%
\bibitem [{\citenamefont {Lin}\ and\ \citenamefont
  {Tol}(2021)}]{lin2021elastic}%
  \BibitemOpen
  \bibfield  {author} {\bibinfo {author} {\bibfnamefont {Z.}~\bibnamefont
  {Lin}}\ and\ \bibinfo {author} {\bibfnamefont {S.}~\bibnamefont {Tol}},\
  }\bibfield  {title} {\enquote {\bibinfo {title} {Elastic metasurfaces for
  full wavefront control and low-frequency energy harvesting},}\ }\href@noop {}
  {\bibfield  {journal} {\bibinfo  {journal} {J. Vib. Acoust.}\ }\textbf
  {\bibinfo {volume} {143}},\ \bibinfo {pages} {061005} (\bibinfo {year}
  {2021})}\BibitemShut {NoStop}%
\bibitem [{\citenamefont {Lan}\ \emph {et~al.}(2017)\citenamefont {Lan},
  \citenamefont {Li}, \citenamefont {Xu},\ and\ \citenamefont
  {Liu}}]{lan2017manipulation}%
  \BibitemOpen
  \bibfield  {author} {\bibinfo {author} {\bibfnamefont {J.}~\bibnamefont
  {Lan}}, \bibinfo {author} {\bibfnamefont {Y.}~\bibnamefont {Li}}, \bibinfo
  {author} {\bibfnamefont {Y.}~\bibnamefont {Xu}}, \ and\ \bibinfo {author}
  {\bibfnamefont {X.}~\bibnamefont {Liu}},\ }\bibfield  {title} {\enquote
  {\bibinfo {title} {Manipulation of acoustic wavefront by gradient metasurface
  based on helmholtz resonators},}\ }\href@noop {} {\bibfield  {journal}
  {\bibinfo  {journal} {Sci. Rep.}\ }\textbf {\bibinfo {volume} {7}},\ \bibinfo
  {pages} {1--9} (\bibinfo {year} {2017})}\BibitemShut {NoStop}%
\bibitem [{\citenamefont {Popa}\ and\ \citenamefont
  {Cummer}(2014)}]{popa2014non}%
  \BibitemOpen
  \bibfield  {author} {\bibinfo {author} {\bibfnamefont {B.~I.}\ \bibnamefont
  {Popa}}\ and\ \bibinfo {author} {\bibfnamefont {S.~A.}\ \bibnamefont
  {Cummer}},\ }\bibfield  {title} {\enquote {\bibinfo {title} {Non-reciprocal
  and highly nonlinear active acoustic metamaterials},}\ }\href@noop {}
  {\bibfield  {journal} {\bibinfo  {journal} {Nat. Commun.}\ }\textbf {\bibinfo
  {volume} {5}},\ \bibinfo {pages} {1--5} (\bibinfo {year} {2014})}\BibitemShut
  {NoStop}%
\bibitem [{\citenamefont {Guo}\ \emph {et~al.}(2018)\citenamefont {Guo},
  \citenamefont {Gusev}, \citenamefont {Bertoldi},\ and\ \citenamefont
  {Tournat}}]{guo2018manipulating}%
  \BibitemOpen
  \bibfield  {author} {\bibinfo {author} {\bibfnamefont {X.}~\bibnamefont
  {Guo}}, \bibinfo {author} {\bibfnamefont {V.~E.}\ \bibnamefont {Gusev}},
  \bibinfo {author} {\bibfnamefont {K.}~\bibnamefont {Bertoldi}}, \ and\
  \bibinfo {author} {\bibfnamefont {V.}~\bibnamefont {Tournat}},\ }\bibfield
  {title} {\enquote {\bibinfo {title} {Manipulating acoustic wave reflection by
  a nonlinear elastic metasurface},}\ }\href@noop {} {\bibfield  {journal}
  {\bibinfo  {journal} {J. Appl. Phys.}\ }\textbf {\bibinfo {volume} {123}},\
  \bibinfo {pages} {124901} (\bibinfo {year} {2018})}\BibitemShut {NoStop}%
\bibitem [{\citenamefont {Guo}\ \emph {et~al.}(2019)\citenamefont {Guo},
  \citenamefont {Gusev}, \citenamefont {Tournat}, \citenamefont {Deng},\ and\
  \citenamefont {Bertoldi}}]{guo2019frequency}%
  \BibitemOpen
  \bibfield  {author} {\bibinfo {author} {\bibfnamefont {X.}~\bibnamefont
  {Guo}}, \bibinfo {author} {\bibfnamefont {V.~E.}\ \bibnamefont {Gusev}},
  \bibinfo {author} {\bibfnamefont {V.}~\bibnamefont {Tournat}}, \bibinfo
  {author} {\bibfnamefont {B.}~\bibnamefont {Deng}}, \ and\ \bibinfo {author}
  {\bibfnamefont {K.}~\bibnamefont {Bertoldi}},\ }\bibfield  {title} {\enquote
  {\bibinfo {title} {Frequency-doubling effect in acoustic reflection by a
  nonlinear, architected rotating-square metasurface},}\ }\href@noop {}
  {\bibfield  {journal} {\bibinfo  {journal} {Phys. Rev. E}\ }\textbf {\bibinfo
  {volume} {99}},\ \bibinfo {pages} {052209} (\bibinfo {year}
  {2019})}\BibitemShut {NoStop}%
\bibitem [{\citenamefont {Ypma}(1995)}]{ypma1995historical}%
  \BibitemOpen
  \bibfield  {author} {\bibinfo {author} {\bibfnamefont {T.~J.}\ \bibnamefont
  {Ypma}},\ }\bibfield  {title} {\enquote {\bibinfo {title} {Historical
  development of the newton--raphson method},}\ }\href@noop {} {\bibfield
  {journal} {\bibinfo  {journal} {SIAM Rev.}\ }\textbf {\bibinfo {volume}
  {37}},\ \bibinfo {pages} {531--551} (\bibinfo {year} {1995})}\BibitemShut
  {NoStop}%
\bibitem [{\citenamefont {Zhu}\ \emph {et~al.}(2015)\citenamefont {Zhu},
  \citenamefont {Zou}, \citenamefont {Li}, \citenamefont {Jiang}, \citenamefont
  {Tu}, \citenamefont {Liang},\ and\ \citenamefont
  {Cheng}}]{zhu2015dispersionless}%
  \BibitemOpen
  \bibfield  {author} {\bibinfo {author} {\bibfnamefont {Y.~F.}\ \bibnamefont
  {Zhu}}, \bibinfo {author} {\bibfnamefont {X.~Y.}\ \bibnamefont {Zou}},
  \bibinfo {author} {\bibfnamefont {R.~Q.}\ \bibnamefont {Li}}, \bibinfo
  {author} {\bibfnamefont {X.}~\bibnamefont {Jiang}}, \bibinfo {author}
  {\bibfnamefont {J.}~\bibnamefont {Tu}}, \bibinfo {author} {\bibfnamefont
  {B.}~\bibnamefont {Liang}}, \ and\ \bibinfo {author} {\bibfnamefont {J.~C.}\
  \bibnamefont {Cheng}},\ }\bibfield  {title} {\enquote {\bibinfo {title}
  {Dispersionless manipulation of reflected acoustic wavefront by subwavelength
  corrugated surface},}\ }\href@noop {} {\bibfield  {journal} {\bibinfo
  {journal} {Sci. Rep.}\ }\textbf {\bibinfo {volume} {5}},\ \bibinfo {pages}
  {1--12} (\bibinfo {year} {2015})}\BibitemShut {NoStop}%
\bibitem [{\citenamefont {Froehly}\ \emph {et~al.}(2011)\citenamefont
  {Froehly}, \citenamefont {Courvoisier}, \citenamefont {Mathis}, \citenamefont
  {Jacquot}, \citenamefont {Furfaro}, \citenamefont {Giust}, \citenamefont
  {Lacourt},\ and\ \citenamefont {Dudley}}]{froehly2011arbitrary}%
  \BibitemOpen
  \bibfield  {author} {\bibinfo {author} {\bibfnamefont {L.}~\bibnamefont
  {Froehly}}, \bibinfo {author} {\bibfnamefont {F.}~\bibnamefont
  {Courvoisier}}, \bibinfo {author} {\bibfnamefont {A.}~\bibnamefont {Mathis}},
  \bibinfo {author} {\bibfnamefont {M.}~\bibnamefont {Jacquot}}, \bibinfo
  {author} {\bibfnamefont {L.}~\bibnamefont {Furfaro}}, \bibinfo {author}
  {\bibfnamefont {R.}~\bibnamefont {Giust}}, \bibinfo {author} {\bibfnamefont
  {P.}~\bibnamefont {Lacourt}}, \ and\ \bibinfo {author} {\bibfnamefont
  {J.}~\bibnamefont {Dudley}},\ }\bibfield  {title} {\enquote {\bibinfo {title}
  {Arbitrary accelerating micron-scale caustic beams in two and three
  dimensions},}\ }\href@noop {} {\bibfield  {journal} {\bibinfo  {journal}
  {Opt. Express}\ }\textbf {\bibinfo {volume} {19}},\ \bibinfo {pages}
  {16455--16465} (\bibinfo {year} {2011})}\BibitemShut {NoStop}%
\bibitem [{\citenamefont {Zhang}\ \emph {et~al.}(2014)\citenamefont {Zhang},
  \citenamefont {Li}, \citenamefont {Zhu}, \citenamefont {Zhu}, \citenamefont
  {Yang}, \citenamefont {Wang}, \citenamefont {Yin},\ and\ \citenamefont
  {Zhang}}]{zhang2014generation}%
  \BibitemOpen
  \bibfield  {author} {\bibinfo {author} {\bibfnamefont {P.}~\bibnamefont
  {Zhang}}, \bibinfo {author} {\bibfnamefont {T.}~\bibnamefont {Li}}, \bibinfo
  {author} {\bibfnamefont {J.}~\bibnamefont {Zhu}}, \bibinfo {author}
  {\bibfnamefont {X.}~\bibnamefont {Zhu}}, \bibinfo {author} {\bibfnamefont
  {S.}~\bibnamefont {Yang}}, \bibinfo {author} {\bibfnamefont {Y.}~\bibnamefont
  {Wang}}, \bibinfo {author} {\bibfnamefont {X.}~\bibnamefont {Yin}}, \ and\
  \bibinfo {author} {\bibfnamefont {X.}~\bibnamefont {Zhang}},\ }\bibfield
  {title} {\enquote {\bibinfo {title} {Generation of acoustic self-bending and
  bottle beams by phase engineering},}\ }\href@noop {} {\bibfield  {journal}
  {\bibinfo  {journal} {Nat. Commun.}\ }\textbf {\bibinfo {volume} {5}},\
  \bibinfo {pages} {1--9} (\bibinfo {year} {2014})}\BibitemShut {NoStop}%
\bibitem [{\citenamefont {Sklan}(2015)}]{sklan2015splash}%
  \BibitemOpen
  \bibfield  {author} {\bibinfo {author} {\bibfnamefont {S.~R.}\ \bibnamefont
  {Sklan}},\ }\bibfield  {title} {\enquote {\bibinfo {title} {Splash, pop,
  sizzle: Information processing with phononic computing},}\ }\href@noop {}
  {\bibfield  {journal} {\bibinfo  {journal} {AIP Adv.}\ }\textbf {\bibinfo
  {volume} {5}},\ \bibinfo {pages} {053302} (\bibinfo {year}
  {2015})}\BibitemShut {NoStop}%
\bibitem [{\citenamefont {Rudenko}(2006)}]{rudenko2006giant}%
  \BibitemOpen
  \bibfield  {author} {\bibinfo {author} {\bibfnamefont {O.~V.}\ \bibnamefont
  {Rudenko}},\ }\bibfield  {title} {\enquote {\bibinfo {title} {Giant
  nonlinearities in structurally inhomogeneous media and the fundamentals of
  nonlinear acoustic diagnostic techniques},}\ }\href@noop {} {\bibfield
  {journal} {\bibinfo  {journal} {Phys.-Uspekhi}\ }\textbf {\bibinfo {volume}
  {49}},\ \bibinfo {pages} {69} (\bibinfo {year} {2006})}\BibitemShut {NoStop}%
\end{thebibliography}%

\end{document}